\documentclass[useAMS,usenatbib]{mn2e}
\usepackage{txfonts}
\usepackage{graphics}
\usepackage{times}

\def\del#1{{}}

\newcommand{\msun}{\rm\,M_\odot}
\newcommand{\hompc}{\,h\,{\rm Mpc}^{-1}}
\newcommand{\mpcoh}{\,h^{-1}\,{\rm Mpc}}
\newcommand{\vecx}{\bmath{x}}
\newcommand{\vecv}{\bupsilon}
\newcommand{\cmatrix}{\mathbfss{C}}
\newcommand{\mmatrix}{\mathbfss{M}}
\renewcommand{\div}{\mathrm{div}}

\title[Peculiar velocities and evolution-bias]{Galaxy peculiar velocities and evolution-bias}
\author[W.J. Percival and B.M. Sch\"{a}fer]
{Will J. Percival\thanks{E-mail: will.percival@port.ac.uk} and Bj{\"o}rn Malte Sch\"{a}fer\\
Institute of Cosmology and Gravitation, University of Portsmouth, Mercantile House, Hampshire Terrace, Portsmouth P01 2EG, United Kingdom}

\begin{document}
\pagerange{\pageref{firstpage}--\pageref{lastpage}}
\pubyear{2007}
\maketitle
\label{firstpage}

\begin{abstract}
  Galaxy bias can be split into two components: a formation-bias based
  on the locations of galaxy creation, and an evolution-bias that
  details their subsequent evolution. In this letter we consider
  evolution-bias in the peaks model. In this model, galaxy formation
  takes place at local maxima in the density field, and we analyse the
  subsequent peculiar motion of these galaxies in a linear model of
  structure formation. The peak restriction yields differences in the
  velocity distribution and correlation between the galaxy and the
  dark matter fields, which causes the evolution-bias component of the
  total bias to evolve in a scale-dependent way. This mechanism
  naturally gives rise to a change in shape between galaxy and matter
  correlation functions that depends on the mean age of the galaxy
  population. This model predicts that older galaxies would be more
  strongly biased on large scales compared to younger galaxies. Our
  arguments are supported by a Monte-Carlo simulation of galaxy pairs
  propagated using the Zel'dovich-approximation for describing
  linear peculiar galaxy motion.
\end{abstract}


\section{introduction}
A number of interesting observational results have recently been
published related to the bias of galaxies. On large scales
$k<0.2\hompc$, the clustering of the 2dFGRS \citep{colless03} and SDSS
\citep{york00} main galaxies is significantly different
\citep{cole05,percival07b,sanchez07}: There is an excess of clustering
for the SDSS main sample galaxies on scales greater than
$k<0.2\hompc$, which is not seen in the 2dFGRS galaxies. If
uncorrected, this impacts on cosmological parameter measurements
utilising the power spectrum shape. If the SDSS main galaxy sample is
split by luminosity, we see that the bias depends on the $r$-band
galaxy luminosity \citep{tegmark04, percival07b}; and a similar trend
was observed in the 2dFGRS sample \citep{norberg01}. If the SDSS
galaxies are split in both colour and luminosity, red galaxies are
more clustered than blue galaxies, but this is not a simple trend
\citep{swanson07}. Red galaxies have the strongest relative bias on
large scales independent of luminosity, while the bias of blue
galaxies appears to rise monotonically with luminosity. In addition,
there appear to be a number of cosmic conspiracies when considering
the cosmological evolution of clustering of different galaxy
classes. The amplitude of the clustering of the SDSS LRGs remains
roughly constant with redshift \citep{eisenstein05, tegmark06}. In
this case, the growth over time of the matter power spectrum is
approximately canceled by a drop in the average bias. \citet{croom05}
considered the evolution of the clustering amplitude of quasars in the
2dFQRS: They found that the amplitude of quasar clustering increases
with increasing redshift. If quasars Poisson sample the matter
distribution in the Universe, however, then we would rather expect
their large-scale clustering amplitude to be proportional to the
linear growth factor, decreasing with increasing redshift. There is
therefore still a great deal still to understand about galaxy bias.

Models of galaxy bias depend on two different processes, in addition
to the clustering of the underlying matter field: formation-bias
describes the effect of galaxies forming at local extrema in the
density field. These galaxies then move following the large-scale
matter velocity field, resulting in evolution-bias. Although the
galaxies may move locally with the matter velocity field, their
peculiar velocity {\em distribution} does not have to match that of
the matter. Ultimately, evolution-bias will lead galaxies to merge
together, reducing their number density and changing the clustering
properties of the population as a whole. In the simplest model of
structure growth, galaxies are always hosted within a halo whose mass
increases with time. In such a picture, evolution-bias is matched by
the increase in the average halo mass which hosts a particular galaxy,
after it has formed. A stronger evolution-bias growth indicates an
increased likelihood of rapid halo mass growth through a continual
merging process. In order to model the bias of an observed galaxy
population we also need to include the observational selection
function of the sample, which will include the galaxy formation time
distributions.

Of the decomposition of galaxy bias into the two components, it is
expected that the formation-bias is the largest contributor for almost
all galaxy populations: the peaks formalism \citep{bbks} predicts a
strong clustering amplitude for galaxy formation at high redshift,
which takes the matter field a long time to match. Formation-bias
evolves strongly with redshift so, for a population of similar
galaxies observed at different redshifts, this will dominate the
change in the bias with time. This is different from evolution-bias,
which gives the change in the bias of the same set of galaxies.

Driven by the observational results discussed above, we now reconsider
evolution-bias from a theoretical point of view. Although galaxy
formation itself is a non-linear process, the velocity field of
galaxies on large scales can be sufficiently well described using
linear theory. Previous work on evolution-bias has assumed a local
model where the distribution of galaxy velocities matches that of the
density field \citep{fry96, tegmark98}. This simplifies the relevant
formulae, and an analytic description of evolution-bias is achievable
(this method is reviewed in Sect.~\ref{sec:bias_evol}). However, if
galaxies formed at the peaks of the smoothed density field, the
expected {\em distributions} of the galaxy peculiar velocities and of
the matter field will not match \citep{bbks}. This effect is due to
the nonzero covariance between the velocity field and the density
gradient and complicates the evolution-bias scenario introduced by
\citet{fry96}. We outline the theory of this approach in
Sects.~\ref{sec:growth_pair} and~\ref{sec:vbias}, describe the
Monte-Carlo sampling technique we use to follow evolution-bias in
Sect.~\ref{sec:mc_peak} and summarise in Sect.~\ref{sec:summary}.

Throughout, the cosmological model assumed is the standard
$\Lambda$CDM cosmology with adiabatic initial perturbations. Choices
for the relevant parameter values are: $\Omega_m=0.25$,
$\Omega_\Lambda=0.75$,
$H_0=100\,h\,\mbox{km~}\mbox{s}^{-1}\mbox{Mpc}^{-1}$ with $h=0.72$,
$\Omega_b=0.04$, $n_s=1$ and $\sigma_8=0.8$. For the matter power
spectrum we make the ansatz $P(k)\propto k^{n_s}T^2(k)$ with the
transfer $T(k)$ function given by \citet{bbks} and the shape parameter
given by \citep{1995ApJS..100..281S}. Where necessary, we impose a
Gaussian smoothing on the density field with the filter scale
$r_\mathrm{scale}$, which is related to the mass scale
$M_\mathrm{scale}=4\pi/3\Omega_m\rho_\mathrm{crit}r_\mathrm{scale}^3$
of the objects considered.

\section{Analytic models of evolution-bias}\label{sec:bias_evol}

\subsection{Limit of small overdensities}

Following the derivation of \citet{fry96}, we assume that the peculiar
velocities of galaxies match those of the matter field at the same
locations, $\vecv$. The continuity equation for the galaxy density can
be written $\partial_t\rho_{\rm gal}=-\div(\rho_{\rm gal}\vecv)$, and
for the matter $\partial_t{\rho}_{\rm CDM}=-\div(\rho_{\rm
  CDM}\vecv)$. By defining $\delta\equiv\rho/\bar{\rho}-1$, the
continuity equation for the matter reduces to $\partial_t{\delta}_{\rm
  CDM}(a)=-\div\vecv(a)$, for small $\delta_{\rm CDM}$. Solving the
system consisting of the continuity and Euler-equations yields the
result that the evolution of $\delta_{\rm CDM}$ is driven by the
standard linear growth factor $D(a)$, so that $\delta_{\rm
  CDM}(a)=D(a)\delta_{\rm CDM}(1)$, with $D(1)=1$ at present day.

If $(\delta_{\rm gal}-\delta_{\rm CDM})$ is small everywhere, then
$\delta_{\rm gal}(a)$ satisfies the same linear continuity equation as
$\delta_{\rm CDM}(a)$, giving $\dot{\delta}_{\rm
  gal}(a)=-\div\vecv(a)$. In this limit,
\begin{equation}
\delta_{\rm gal}(a)=\delta_{\rm gal}(1)+(D(a)-1)\delta_{\rm CDM}(1).
\label{eq:delta_evol}
\end{equation}
Note that the assumption of small $(\delta_{\rm gal}-\delta_{\rm
  CDM})$ is important, because the densities of the dark matter and
galaxy fields need to be matched in the full continuity equations, in
order for the expected growth in $\delta_{\rm gal}$ to be equal to
that of $\delta_{\rm CDM}$. This assumption is equivalent to stating
that the distribution of the peculiar velocities of the galaxies is
the same as that of the matter field.

\subsection{Starting from linear bias}

This idea has been extended by \citet{fry96}, who considered a galaxy
overdensity field with a linear bias at present day $\delta_{\rm
  gal}(1)=b(1)\delta_{\rm CDM}(1)$. Substituting this into
eqn.~(\ref{eq:delta_evol}) gives $\delta_{\rm
  gal}(a)=[D(a)-1+b(1)]\delta_{\rm CDM}(1)$. In this model the linear
bias is is related to the growth function by
\begin{equation} 
b(a) = \frac{D(a)-1+b(1)}{D(a)}.
\label{eq:b_evol}
\end{equation}
Furthermore \citet{tegmark98} considered the evolution-bias of
galaxies following the more general stochastic bias model of
\citet{dekel99} and derive cross correlations between the matter
density and galaxy fields. To simplify the analysis we follow
\citet{tegmark98} and define the vector
\begin{equation}
  {\vecx}(a)=\left(\begin{array}{c}
    \delta_{\rm CDM}(a) \\
    \delta_{\rm gal}(a) \end{array}
  \right).
\end{equation}
Forcing the galaxy overdensity field to obey the bias model of
\citet{dekel99} at present day allows the galaxy and dark matter
distributions to be related by an additional correlation coefficient
$r$. At present day, the covariance matrix can be written
\begin{equation}
  \cmatrix(1)\equiv\langle{\vecx}(1){\vecx}^t(1)\rangle
  =\langle \delta_{\rm CDM}^2(1) \rangle
  \left(\begin{array}{cc} 
    1      & b(1)r(1) \\
    b(1)r(1) & b(1)^2 \end{array}
  \right).
\end{equation}
In this notation, the continuity equation for the galaxy overdensity
field coupled with the linear growth factor for the matter gives
\begin{equation}
  {\vecx}(a) = \mmatrix(a){\vecx}(1)\,\,{\rm,~where}\,\,
  \mmatrix\equiv\left(\begin{array}{cc} 
    D(a)   & 0 \\
    D(a)-1 & 1 \end{array}
  \right).
\end{equation}
The corresponding evolution of the covariance matrix
$\cmatrix(a)=\mmatrix(a)\cmatrix(1)\mmatrix^t(a)$ shows that the
evolution of the galaxy overdensity field can be rewritten in terms of
$r(a)$ and $b(a)$ \citep{tegmark98}, and that the bias model does not
change its form. For $r=1$, the formula reduces to the relation of
\citet{fry96}.

\subsection{Bias for Gaussian random fields}

In work predating \citet{fry96} and \citet{tegmark98}, \citet{bbks}
considered the evolution of bias for galaxies that form at peaks of
the density field after smoothing on a certain mass scale. On large
scales, the peak density field can be considered as a continuous
Gaussian random field itself, statistically independent and
uncorrelated with the background, which drives the dynamics of the
galaxies. The galaxy and matter density field evolve independently due
to the peak-background split, as independent $k$-modes are important
for their respective evolution.

Using this model \citet{bbks} showed that the large-scale bias
evolves as $b(a)\propto\delta_c/\sigma_M(a)+1$ for a smoothed Gaussian
density field with variance $\sigma_M(a)$, where $\delta_c$ is the
overdensity threshold above which galaxies form. On linear scales, the
variance of the density field $\sigma_M(a)$ is proportional to the
linear growth factor $\sigma_M(a)\propto D(a)$. Consequently, if we
normalise the bias at present day, we recover eqn.~(\ref{eq:b_evol})
for the evolution of the bias. A key assumption made in the derivation
is the treatment of the galaxy distribution as a continuous field and
that the evolution in the comoving number density of peaks is driven
by the evolution of the background density, which is in effect a
reformulation of the continuity equation for the galaxy field: Forcing
the number density of galaxies to evolve with the background density
is akin to the assumption of small $(\delta_{\rm gal}-\delta_{\rm
  CDM})$ that led to the linearised continuity equation for the galaxy
field.

\section{structure growth from pair velocities}  
\label{sec:growth_pair}

The above descriptions of evolution-bias can also be understood in
terms of the peculiar velocities of a discrete set of galaxies. The
variance of the peculiar velocity distribution for each galaxy is
\begin{equation}
  \sigma_\upsilon^2=\frac{[aH(z)f]^2}{2\pi^2}\int_0^\infty dk\,P(k),
  \label{eq:sigv_mass}
\end{equation}
where $H(z)$ is the Hubble parameter and $f\equiv d\log D/d\log a$.
If we assume that $\delta$ is small everywhere, the expected infall
velocity is zero for all pairs of galaxies: we are as likely to choose
galaxies that are moving apart (in comoving space), as peaks moving
together. If galaxy peculiar velocities were uncorrelated, the galaxy
density field would not develop clustering as any initial clustering
pattern will be destroyed by the random diffusion. This is not the
case for velocity fields driven by correlated Gaussian random
overdensity fields as the velocities of galaxy pairs of separation $r$
are correlated according to
\begin{equation}
  C_\upsilon(r) = \langle {\vecv}_1\cdot{\vecv}_2 \rangle
  = \frac{[aH(z)f]^2}{2\pi^2}\int_0^\infty dk\,P(k)j_0(kr).
  \label{eq:xiv_mass}
\end{equation}
Velocities of pairs of galaxies will be strongly correlated when $r$
is small as both galaxies preferentially move in the same direction,
being part of the same bulk flow, but velocities of galaxy pairs with
large separations are less strongly correlated. This difference means
that there is a net influx of galaxy pairs from large to small
separations, which leads to the growth in the clustering strength. For
a pair of galaxies of separation $r$, the variance of the peculiar
velocity difference is
\begin{equation}
  \langle |{\vecv}_1-{\vecv}_2|^2 \rangle = 2[\sigma_\upsilon^2-C_\upsilon(r)],
  \label{eq:vdiff_mass}
\end{equation}
where $\sigma_\upsilon^2$ is a diffusion term.

Changing the selection criteria of galaxy pair-velocities naturally
changes the evolution of galaxy clustering. An extreme example of a
galaxy field with an evolution-bias would be the assumption that
galaxies only form at locations where the peculiar velocity
${\vecv}=0$. After formation, galaxies at these locations do not move,
and the galaxy overdensity and clustering properties are constant in
comoving space. In contrast to the model proposed by \citet{bbks},
this is not physically motivated, because there is no a-priori reason
to link places in the velocity field where ${\vecv}\simeq0$ with
galaxy formation.  The derivation leading to
eqn.~(\ref{eq:delta_evol}) breaks down for this model because it
relied on matching the galaxy and matter overdensity fields, which is
clearly broken. In this situation, the galaxies still locally ``move
with the matter'' although the {\em distribution} of galaxy velocities
does not match the {\em distribution} of velocities of the dark
matter.

\section{evolution-bias in the peaks formalism} \label{sec:vbias}

We now consider the evolution of a set of galaxies that form at the
peaks of a smoothed density field, where there is the most rapid
increase in density. After formation, the galaxies leave the peaks,
whose positions are fixed in comoving space, and move with the matter
flow.

\begin{figure}
\centering
\resizebox{0.9\columnwidth}{!}{\includegraphics{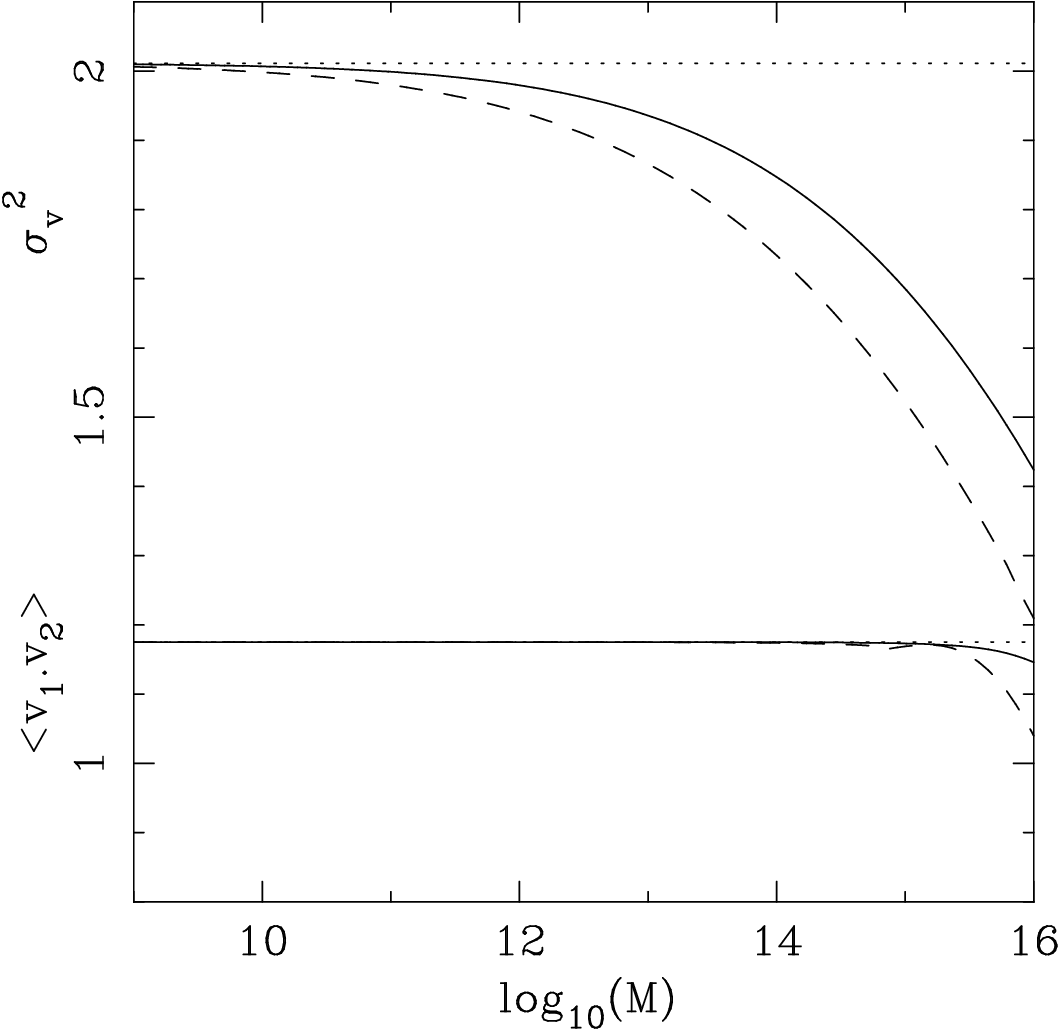}}
\caption{The expected variance for velocities in a Gaussian random
  field smoothed with a Gaussian filter (solid lines), and for the
  unsmoothed field (dotted lines), plotted against smoothing scale
  parametrised by the mass enclosed within the filter (upper lines).
  The expected velocity correlation between two points separated by
  $200\mpcoh$, under the same assumptions are shown by the lower
  lines. The decrease in velocity variance caused by additionally
  selecting sampling points with vanishing gradient in the smoothed
  density field is indicated by the dashed lines.}
\label{fig:sig_lM}
\end{figure}

We wish to estimate the average velocity of a matter concentration at
a peak in the field, so we will need to consider the density and
velocity fields after smoothing by a filter of width $r_{\rm
  smooth}$. The velocity dispersion and correlation can be calculated
by replacing $P(k)$ with the smoothed power spectrum $\bar{P}(k)$ in
eqns.~(\ref{eq:sigv_mass}), (\ref{eq:xiv_mass}) and
(\ref{eq:vdiff_mass}). For large separations $r\gg r_{\rm smooth}$,
the Bessel function in eqn~(\ref{eq:xiv_mass}) takes precedence over
the smoothing of the field, and $C_\upsilon(r)$ tends towards the
value for the unsmoothed field. In contrast, the variance of the
velocity for each galaxy $\sigma_\upsilon^2$
(eqn.~\ref{eq:sigv_mass}), which is independent of $r$, does depend on
the smoothing applied to the field.  The effects of the smoothing
scale on $\sigma_\upsilon^2$ and $C_\upsilon(r)$ are shown in
Fig.~(\ref{fig:sig_lM}), for a pair separation of $200\mpcoh$.

In order to select peaks in the smoothed density field, we need to set
two further constraints on the galaxy locations: The density gradient
has to be zero, and the local curvature needs to be positive
definite. Because the velocities are correlated with the density
gradient, placing galaxies at locations where the gradient is zero
reduces the velocity variance $\sigma_\upsilon^2$ from that of
eqn~(\ref{eq:sigv_mass}) by a factor $(1-\gamma_\upsilon^2)$
\citep{bbks,szalay87,peacock87}, where
\begin{equation}
\gamma_\upsilon = \frac{\sigma_0^2}{\sigma_{-1}\sigma_{1}},\quad\mathrm{and}\quad
\sigma_j^2=4\pi\int_0^\infty\,dk\,k^{2j+2}P(k).
\label{eq:sigj}
\end{equation}
For a smoothed field, $P(k)$ needs to be replaced by $\bar{P}(k)$ with
the consequence that $\gamma_\upsilon\to1$ as $r_{\rm
  smooth}\to\infty$. In the limit of large $r$, the velocity
correlation function is unchanged by the peak constraint: In this
limit, velocity-velocity correlations dominate over velocity-gradient,
or gradient-gradient correlations because of the additional $k^{-1}$
terms in the velocity-dependent integrands. The effect of including
both the smoothing of the field and the zero-gradient selection
criteria on $\sigma_\upsilon^2$ and $C_\upsilon(r)$ is shown in
Fig.~(\ref{fig:sig_lM}), for a pair separation of $200\mpcoh$.

The reduction of the dispersion element of eqn.~(\ref{eq:vdiff_mass}),
while keeping the growth component fixed will lead to a
scale-dependent increase in the 2-point galaxy correlation function
(\citet{regos95} also comment on this). From this it is clear that
propagating galaxies along their velocity vectors does not preserve
the shape of the correlation function, and the bias evolution
formalism developed in Sect.~\ref{sec:bias_evol} breaks down.

\section{Monte-Carlo simulation of galaxy pair
  properties} \label{sec:mc_peak}

We use the formalism introduced by \citet{regos95} to set up a
Gaussian random process for determining velocity variances and
correlations in the large-scale structure. \citet{regos95} show how a
$26\times26$ covariance matrix can be constructed for the
multi-variate Gaussian distribution of the properties (overdensity, 3
gradient components, 6 curvature components and 3 velocity components)
of pairs of points in a smoothed Gaussian random field. The matrix
depends on the power spectrum moments given in eqn.~(\ref{eq:sigj})
and from the functions of the pair separation $r$ given by
\begin{equation}
  K_{\ell m} = 4\pi\int_0^\infty\,dk\,k^m j_l(kr) \bar{P}(k).
  \label{eq:K}
\end{equation}
Using these covariance matrices, we can randomly draw realisations of
the properties of pairs of points in the field, assuming they follow a
multi-variate Gaussian distribution. We can also add constraints on
the pairs selected, such as that they must both be at a peak in the
field.

\begin{figure}
  \centering
  \resizebox{0.9\columnwidth}{!}{\includegraphics{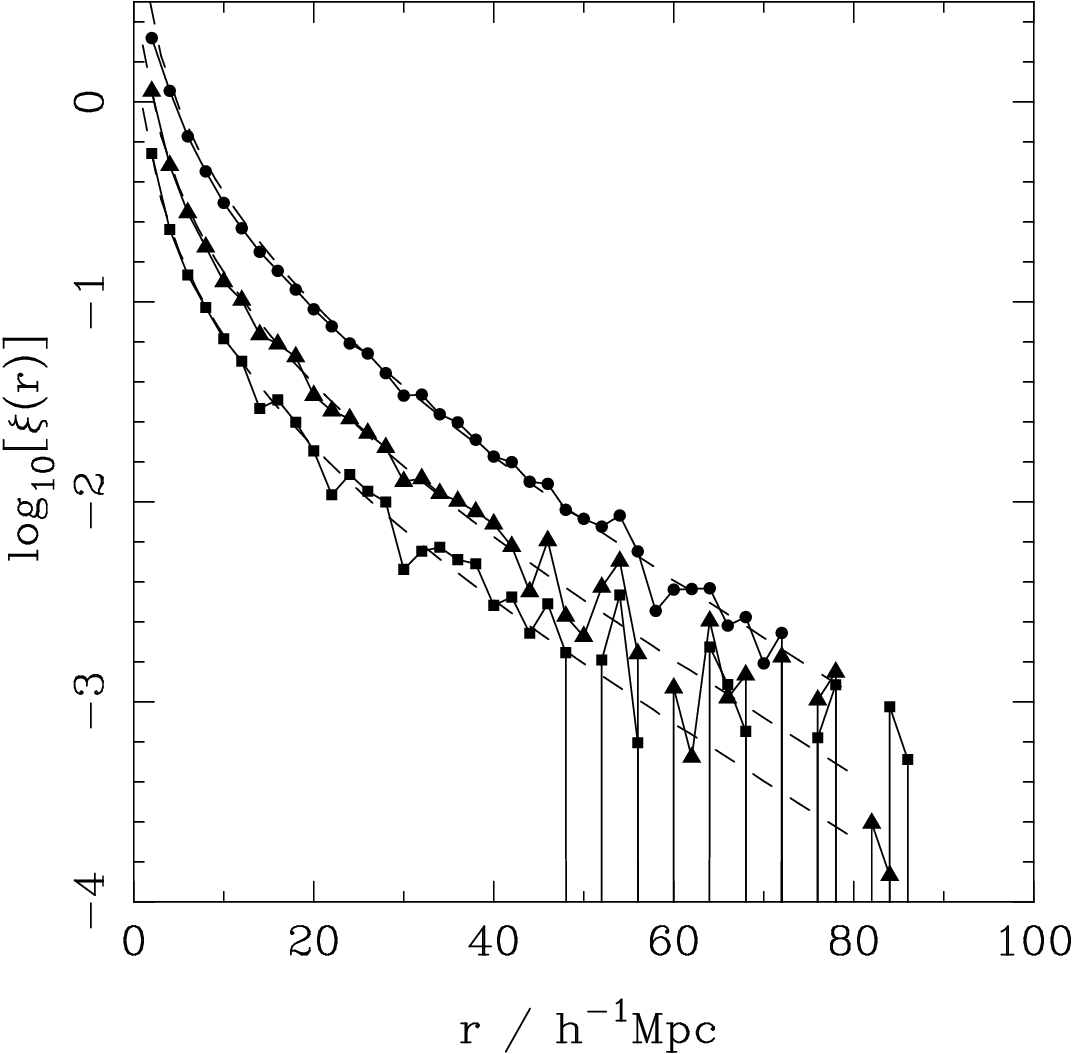}}
  \caption{The correlation function at $z=0$ (circles), $z=1$
    (triangles) and $z=2$ (squares) calculated from a Monte-Carlo
    realisation of $10^8$ independent pairs of points in a Gaussian
    random field. These are compared with the expected correlations
    for linear growth (dashed lines). On large scales, $\xi(r)$
    becomes negative because we only sample pairs of galaxies with
    initial separation $<100\mpcoh$. \label{fig:xi_mass}}
\end{figure}

We first demonstrate that this procedure can reproduce the linear
growth expected for the matter field. We have produced a Monte-Carlo
realisation of $10^8$ pairs of galaxies using the procedure described
above, without any additional constraints on pair selection. This
corresponds to choosing random locations in a matter field. The
peculiar velocities were then used to estimate the motion of galaxies
in linear structure formation using the Zel'dovich approximation. This
is applicable for small displacements, and should therefore accurately
describe the initial effect of evolution-bias after galaxy
formation. The dynamical model is affected by the properties of the
particular dark energy model through the growth function $D(a)$, which
affects the peculiar velocities. From these data, we have calculated
$\xi(r)$ for $z=0,1,2$, assuming that the galaxy pairs evolve from an
initial unclustered comoving distribution (as expected for the matter
in the Universe, which is initially homogeneous). The correlation
functions are plotted in Fig.~\ref{fig:xi_mass} showing excellent
agreement with the expected linear evolution.

We now consider galaxies that form at peaks in this field. The
additional criteria applied within our Monte-Carlo procedure to ensure
that we select peaks are
\begin{enumerate}
\item overdensity threshold $\delta>\delta_c=1.69$,
\item gradient ${\bf \nabla}\delta = 0$,
\item positive definite curvature matrix.
\end{enumerate}
Peaks are rare, so simply selecting peaks from pairs chosen at random
is computationally unfeasible. Instead, where possible, we force the
properties that are required (for example, we only sample from the
tail of a Gaussian distribution to select $\delta>\delta_c$), and then
weight the chosen pairs by the likelihood of making that selection in
a full multi-variate analysis. This gives a weighted distribution with
the same statistical properties as selecting a true Monte-Carlo
sample, but is computationally faster for determining the properties
of the distribution.

\begin{figure}
  \centering
  \resizebox{0.9\columnwidth}{!}{\includegraphics{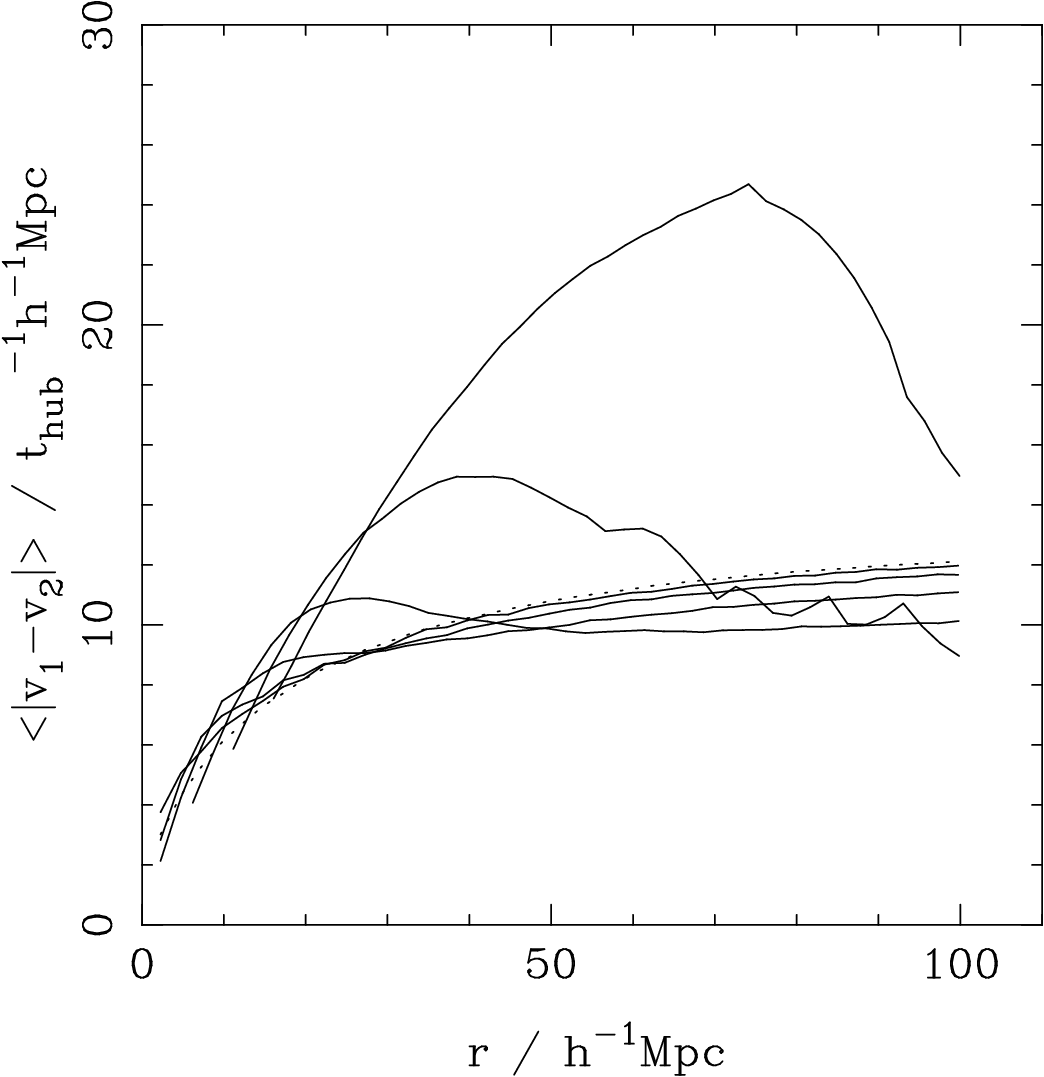}}
  \caption{The expected amplitude of the pair-velocity as a function
    of pair-separation for peaks selected for halos of mass
    $10^{10\ldots 15}\msun$ (solid lines). Increasing the halo mass
    leads to larger infall velocities. These curves were calculated
    from $10^7$ independent pairs of peaks, although each was weighted
    as described in the text, so these correspond to a much smaller
    effective number of pairs. The expected pair-velocity for
    locations selected at random in the matter distribution is shown
    by the dotted line. \label{fig:peak_vel}}
\end{figure}

The expected amplitude of the pair-velocity
$\langle|{\vecv}_1-{\vecv}_2|\rangle$ for galaxies selected at peaks,
is plotted in Fig.~\ref{fig:peak_vel} for halos of mass $10^{10\ldots
  15}\msun$. The expected pair-velocities differ from those of the
mass, because peaks are more likely to be approaching each other than
moving apart at large separations. The evolution in the correlation
function depends on the derivative of the expected pair-velocity,
which controls the net change of pairs with a given separation (when
the distances travelled are small). As we move to increasing
separation, the peak pair-velocity increases rapidly, but then turns
over and starts to decrease. This will lead to the correlation
function decreasing on small scales, and increasing on large scales.

\begin{figure}
\centering
\resizebox{0.9\columnwidth}{!}{\includegraphics{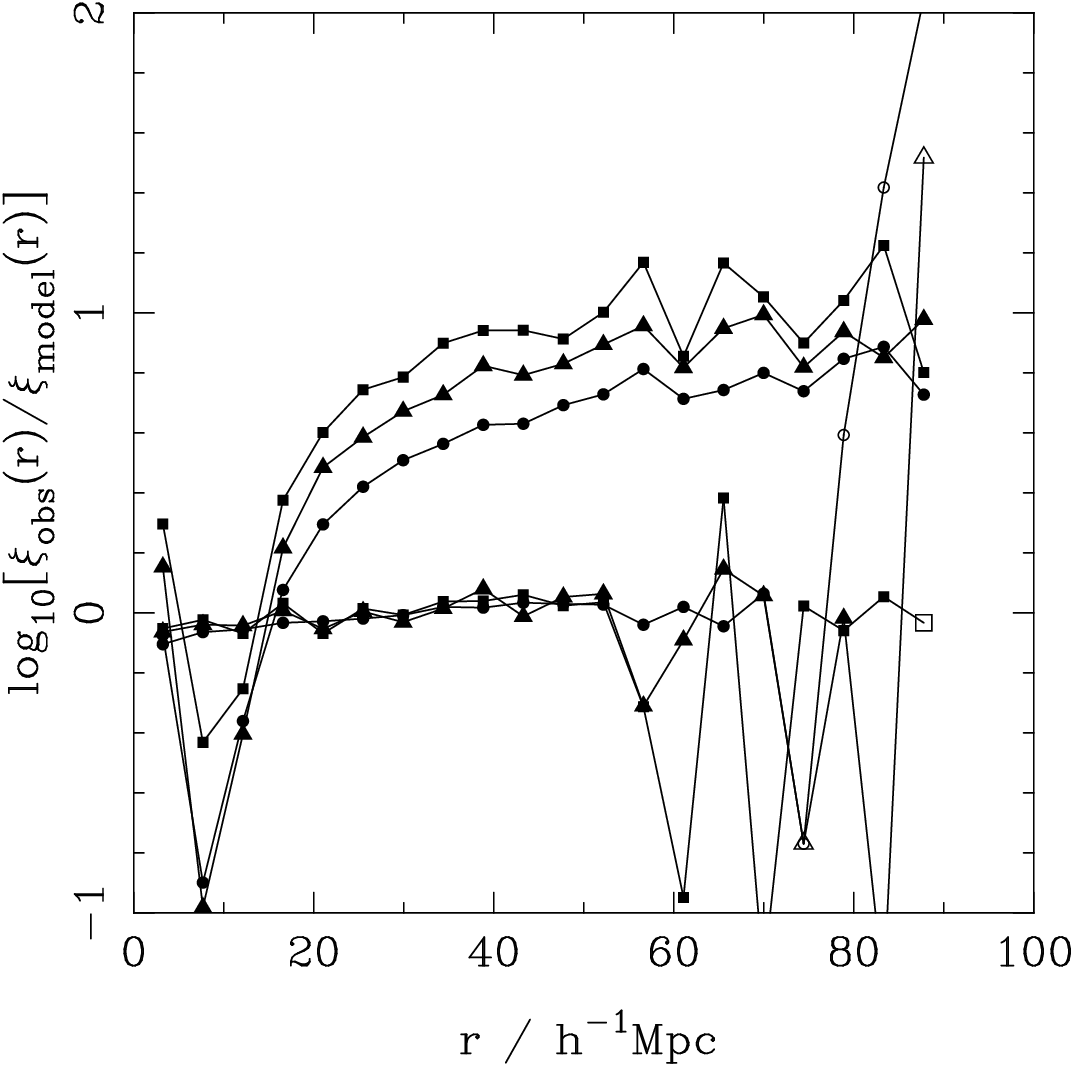}}
\caption{The correlation functions recovered from propagating pairs of
  galaxies along their peculiar velocities, divided by those expected
  from linear evolution of the density field. Correlation functions at
  $z=0$ (circles), $z=1$ (triangles) and $z=2$ (squares), were
  calculated from a Monte-Carlo realisation of $10^8$ independent
  pairs of points in a Gaussian random field (lower lines), and from
  $3\times10^8$ weighted peaks in a Gaussian random field smoothed
  with a Gaussian filter corresponding to a halo mass of
  $10^{12}\msun$ (upper lines).}
\label{fig:xi_ratio}
\end{figure}

We have constructed a weighted Monte-Carlo distribution of
$3\times10^8$ galaxy pairs selected at the peaks in a Gaussian random
field, smoothed by a Gaussian filter with width corresponding to a
halo mass $10^{12}\msun$. In order to demonstrate the effect of peak
selection on the galaxy correlation function, we have followed the
evolution of galaxies pairs from an initially unclustered
distribution. This matches the analysis of pairs of galaxies selected
at random from the matter distribution that led to
Fig.~\ref{fig:xi_mass}, but selects special locations in the velocity
field. The ratios between the recovered galaxy correlation functions
and those expected for the matter at $z=0,1,2$, are shown in
Fig.~\ref{fig:xi_ratio}. As we move from small to large scales, the
bias at any redshift initially decreases, because of the change in
expected infall velocity shown in Fig.~\ref{fig:peak_vel}, but then
increases beyond that of the matter field, as expected from the
analytic arguments discussed in Sect.~\ref{sec:vbias}. Due to the fact
that galaxies only pick up their peculiar velocities after formation,
this plot should not be interpreted as giving the actual galaxy
correlation functions for peaks. To do this, we would need to specify
galaxy formation times and their spatial distribution (or
formation-bias). The relative effect of the evolution-bias on
$\xi(r)$, however, should be the same, and this plot demonstrates that
bias is not a simple function of scale. The large-scale bias is a
decreasing function of time, as the amplitude of the clustering of the
matter field grows to match the strong evolution-bias predicted at
early times.

\section{Summary and discussion}\label{sec:summary}

We have considered a decomposition of galaxy bias into formation-bias
and evolution-bias. The formation- and evolution-biases are
inter-related, with one tending to lead to the other: By choosing
special locations for galaxy formation, we also tend to choose
locations that lead to group evolution that is very different from
that of particles selected at random in the matter field. By analysing
models of evolution-bias, we have argued that this component of bias
has some interesting properties within the peaks model. We have
demonstrated this from both analytic arguments and a Monte-Carlo
procedure based on the work of \citet{regos95}. Evolution-bias
provides a mechanism for producing a scale-dependent bias in 2-pt
correlation measurements, whose importance grows as the galaxies move,
and consequently depends on the time since galaxy formation. Note that
the overall bias of a galaxy sample would, in general, be expected to
be a decreasing function of time because this depends on the
increasing amplitude of the clustering of the matter field.

The inclusion of evolution-bias in a combined model of galaxy bias
would predict that older galaxies have a scale dependent bias with a
correlation function whose shape is less like that of the matter field
than a population of younger galaxies. In the most simple model, the
red luminosity of an individual galaxy increases and then fades with
time. Galaxies can also merge together, forming larger, brighter
objects. The oldest field galaxies tend to be faintest, and should be
most affected by scale-dependent biasing on large scales. Such a
description might explain why the SDSS LRGs and 2dFGRS galaxies seem
to match a simple prescription for galaxy bias, while the SDSS main
galaxies do not \citep{sanchez07}: The SDSS main galaxies sample would
contain these old field galaxies. The evolution-bias considered here
would also impact on measurements of the Baryon Acoustic Oscillation
scale length from the power spectrum or correlation function. Further
simulations, such as provided by semi-analytic techniques, which
combine formation and evolution-bias, are required to consider this
mechanism in more detail and apply it to specific galaxy types. 

\section*{Acknowledgements}
WJP and BMS are grateful for support from STFC.

\bibliography{bibtex/aamnem,bibtex/references}
\bibliographystyle{mn2e}

\appendix

\bsp

\label{lastpage}

\end{document}